\documentclass[10pt,preprint2]{aastex}

\usepackage{natbib}

\shorttitle{HETGS Observation of PZ~Tel}
\shortauthors{Argiroffi et al.}

\received{2004 January 9}
\accepted{2004 February 25}
\begin{document}

\title{High Resolution X-ray Spectroscopy of the Post-T~Tauri Star PZ~Tel} 

\author{C. Argiroffi\altaffilmark{1}, J.J. Drake\altaffilmark{2}, A. Maggio\altaffilmark{3}, G. Peres\altaffilmark{1}, S. Sciortino\altaffilmark{3}, and F.R. Harnden\altaffilmark{2}}

\altaffiltext{1}{Dipartimento di Scienze Fisiche ed Astronomiche, Sezione di Astronomia, Universit\`a di Palermo, Piazza del Parlamento 1, 90134 Palermo, Italy; argi@astropa.unipa.it, peres@astropa.unipa.it.}
\altaffiltext{2}{Smithsonian Astrophysics Observatory, 60 Garden Street, Cambridge, MA 02138; jdrake@cfa.harvard.edu, frh@cfa.harvard.edu.}
\altaffiltext{3}{INAF - Osservatorio Astronomico di Palermo, Piazza del Parlamento 1, 90134 Palermo, Italy; maggio@astropa.unipa.it, sciorti@astropa.unipa.it.}

\begin{abstract} 
We present an analysis of the {\em Chandra} High Energy Transmission Grating Spectrometer observation of the rapidly rotating ($P_{\rm rot}=0.94\,{\rm d}$) post T~Tauri ($\sim20 $\,Myr old) star PZ~Telescopii, in the Tucana association. Using two different methods we have derived the coronal emission measure distribution, $em(T)$, and chemical abundances. The $em(T)$ peaks at $\log T = 6.9$ and exhibits a significant emission measure at temperatures $\log T > 7$. The coronal abundances are generally $\sim 0.5$ times the solar photospheric values that are presumed fairly representative of the composition of the underlying star. A minimum in abundance is seen at a first ionization potential (FIP) of 7-8\,eV, with evidence for higher abundances at both lower and higher FIP, similar to patterns seen in other active stars. From an analysis of the He-like triplet of \ion{Mg}{11} we have estimated electron densities of $\sim 10^{12}-10^{13}\,{\rm cm^{-3}}$. All the coronal properties found for PZ~Tel are much more similar to those of AB~Dor, which is slightly older than PZ~Tel, than to those of the younger T~Tauri star TW~Hya. These results support earlier conclusions that the soft X-ray emission of TW~Hya is likely dominated by accretion activity rather than by a magnetically-heated corona. Our results also suggest that the coronae of pre-main sequence stars rapidly become similar to those of older active main-sequence stars soon after the accretion stage has ended.
\end{abstract} 

\section{INTRODUCTION}
\label{intro}

One of the primary characteristics of low mass pre-main sequence (PMS) stars is their intense X-ray activity. This X-ray emission therefore represents an important means for investigating the properties and evolution of young stellar objects. X-ray activity is present during the evolution of PMS stars both in the initial evolutionary stages of a Classic T~Tauri Star (CTTS, Class I and II sources), during which the star has an accretion disk that surrounds it, and in the subsequent Weak-Line T~Tauri Star (WTTS, Class III sources) phase in which the star has no accreting material and is approaching the zero-age main sequence \citep[ZAMS, ][]{FeigelsonMontmerle1999}. How the presence of accreting material influences the X-ray emission of PMS stars, and for how long, remain questions of debate.

The high resolution X-ray spectra now available with the {\em Chandra} and {\em XMM-Newton} satellites offer the possibility to perform detailed studies of stellar coronae because key emission lines diagnostics can now be resolved. These diagnostics can be used to derive elemental abundances, temperature and density structure of the emitting plasmas.

It is worth noting that very few young stars in star forming regions or associations are sufficiently X-ray bright to allow high resolution X-ray spectroscopy with current instrumentation. In particular, among the CTTSs, \objectname{TW~Hya} is the best studied case to date because it is the nearest ($\sim56$\,pc) known CTTS. TW~Hya shows spectral characteristics very different from those of young but otherwise \emph{normal} active stars \citep{KastnerHuenemoerder2002,StelzerSchmitt2004}: very low plasma temperature ($\log T \sim 6.5$), high density ($\log N_{\rm e} \sim 13$), very low Fe abundance ($A_{\rm Fe}/A_{\rm Fe\sun} \sim 0.2$). \citet{KastnerHuenemoerder2002} and \citet{StelzerSchmitt2004} attributed these characteristics to an accretion shock rather than coronal activity. Without the benefit of high resolution spectra of similar stars for comparison, however, the nature of the peculiarity of TW~Hya remains uncertain.

The situation with slightly more evolved stars is more clear. Studies of \objectname{AB~Dor}, a young active star that has nearly arrived at the ZAMS and which has been observed in the past with several space-borne X-ray observatories, have shown that it is characterized by a hot corona ($T \sim 10^7$\,K) with plasma densities ranging from $6\times10^{10}\,{\rm cm^{-3}}$ at $2\times10^{6}$\,K, to $3\times10^{12}\,{\rm cm^{-3}}$ at $10^{7}$\,K, and a moderately low Fe abundance ($A_{\rm Fe}/A_{\rm Fe\sun} \sim 0.25$, \citealt{MeweKaastra1996,GudelAudard2001}; \citealt*{Sanz-ForcadaMaggio2003}; P. Testa, in preparation;  D. Garc{\'{\i}}a-Alvarez, in preparation). In many respects, AB~Dor can be considered the prototype of very active single stars.

High resolution X-ray spectroscopy of other PMS stars with ages between few $10^{6}$\,yr, typical of CTTSs like TW~Hya, and $10^{8}$\,yr, the ZAMS of solar-type stars like AB~Dor, is crucial for understanding how the characteristics of stellar X-ray activity change during these early evolutionary phases.

In this work, we present a {\em Chandra} High Energy Transmission Grating Spectrometer (HETGS) observation of \objectname{PZ~Telescopii} (HD~174429, HIP~92680). Classified as a K0V star by \citet{Houk1978}, PZ~Tel was determined by \citet{ZuckermanWebb2000} to be a member of the Tucana association, a nearby star forming region about 45\,pc away. From observations with the ROSAT PSPC, \citet{StelzerNeuhauser2000} deduced a PZ~Tel X-ray luminosity of $L_{\rm X}\sim(2.88\pm0.0 8)\times10^{30}\,{\rm erg\,s^{-1}}$. At a Hipparcos parallax distance of $49.7\pm2.9$\,pc \citep{PerrymanLindegren1997}, PZ~Tel is a single star with a rotational period of 0.94\,d (\citealt{CoatesHalprin1980}; \citealt*{InnisCoates1984,InnisThompson1986}). \citet{FavataMicela1998} have estimated a mass of $1.1\,M_{\sun}$ for PZ~Tel and an age of approximately 20\,Myr. Its youth is also confirmed by prominent H$\alpha$ emission and a relatively undepleted Li abundance \citep*{SoderblomKing1998}. Further evidence of its PMS status has been pointed out by \citet{BarnesCollierCameron2000} who deduced from $v \sin i$ and $P_{\rm rot}$ that the minimum radius of PZ~Tel is $R\,\sin i \sim 1.27\,R_{\sun}$: this value is larger than the radius of a main sequence star with the same mass as PZ~Tel.

The analysis of {\em Chandra} HETGS high resolution spectra of PZ~Tel offers us the opportunity to study the coronal properties of a single star which has dissipated its accretion disk and is approaching the ZAMS. This study also allows us to compare the coronal properties of PZ~Tel with those of both younger and older stars, providing a glimpse of the evolution of stellar coronae through the PMS phase. In particular, the CTTSs have disks from which active accretion is still taking place. These stars could have quite different magnetospheric geometries, possibly involving magnetic connections between star and disk \citep[e.g.][]{Montmerle2002}. In the case of coeval and older stars, it appears that coronal activity of PMS stars without an accreting disk can be explained on the same basis as that of main-sequence stars \citep*{FlaccomioMicela2003} --- stellar rotation and convection. In this domain, PZ~Tel provides a new window on phenomena such as the chemical fractionation of elements that is seen to occur in coronae over a wide range of activity level. In the solar case, elements with low first ionization potential (FIP) such as Mg, Fe and Si are seen to be enhanced relative to elements with high FIP, such as O, Ne and Ar \citep[e.g.][]{Feldman1992}. In more active stars, the situation appears somewhat reversed, with elements such as Ne appearing enhanced relative to those with lower FIP \citep[e.g.][and references therein]{Drake2002}. The case of the active but very young post-T Tauri stars remains unexplored at high spectral resolution.

In \S~\ref{obs} and \ref{analysis} we describe the {\em Chandra} observation and  the techniques used in its analysis. Section~\ref{results} presents the resulting coronal temperature structure and abundances. These are discussed in \S~\ref{disc}, in which we also present a comparison of the coronal properties of PZ~Tel with those of AB~Dor and TW~Hya.

\begin{figure*}[t]
\centering
\includegraphics[width=13cm]
{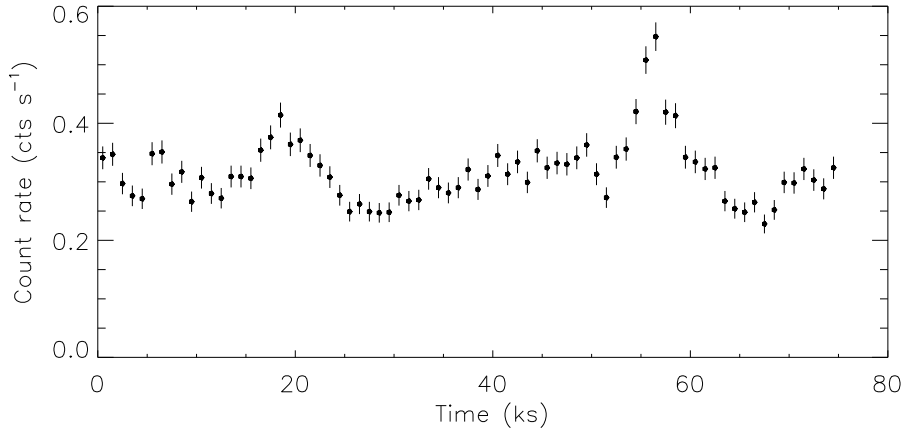}
\caption{Light curve of PZ~Tel obtained from the HEG and MEG spectra (excluding zero-order events) with bin size of 1000\,s.}
\label{fig:pztel_lightcurve}
\end{figure*}

\section{OBSERVATION}
\label{obs}

PZ~Tel was observed with the {\em Chandra} HETGS on 2003 June 7, for 73.9\,ks. The HETGS provides both High Energy (HEG) and Medium Energy Grating (MEG) spectra. The spectral ranges available with these instruments are 1-18\,\AA~(HEG) and 1-25\,\AA~(MEG), with spectral resolutions $\Delta\lambda$ of 0.01\,\AA~and 0.02\,\AA~respectively. We have used CIAO~V3.0 to extract the positive and negative first order spectra for each grating and to compute the effective area of the HETG+ACIS-S combinations. In this latter step, we have also taken into account the attenuation resulting from the build-up of a contamination layer on the ACIS-S filter \citep[e.g.][]{PlucinskySchulz2003} using CIAO Version 1 of the effective area contamination  correction\footnote{http://cxc.harvard.edu/ciao/threads/aciscontam/}. In Fig.~\ref{fig:pztel_lightcurve} we show the light curve of photon arrival times of the HEG and MEG spectra, excluding zeroth order events that are subject to significant pile up. Figure~\ref{fig:pztel_lightcurve} shows the X-ray flux to be characterized by variability on time scales of a few to tens of ks, with two flare-like events occurring at about 20\,ks and 55\,ks after the start of the observation.

\begin{figure*}
\centering
\includegraphics[width=15cm]
{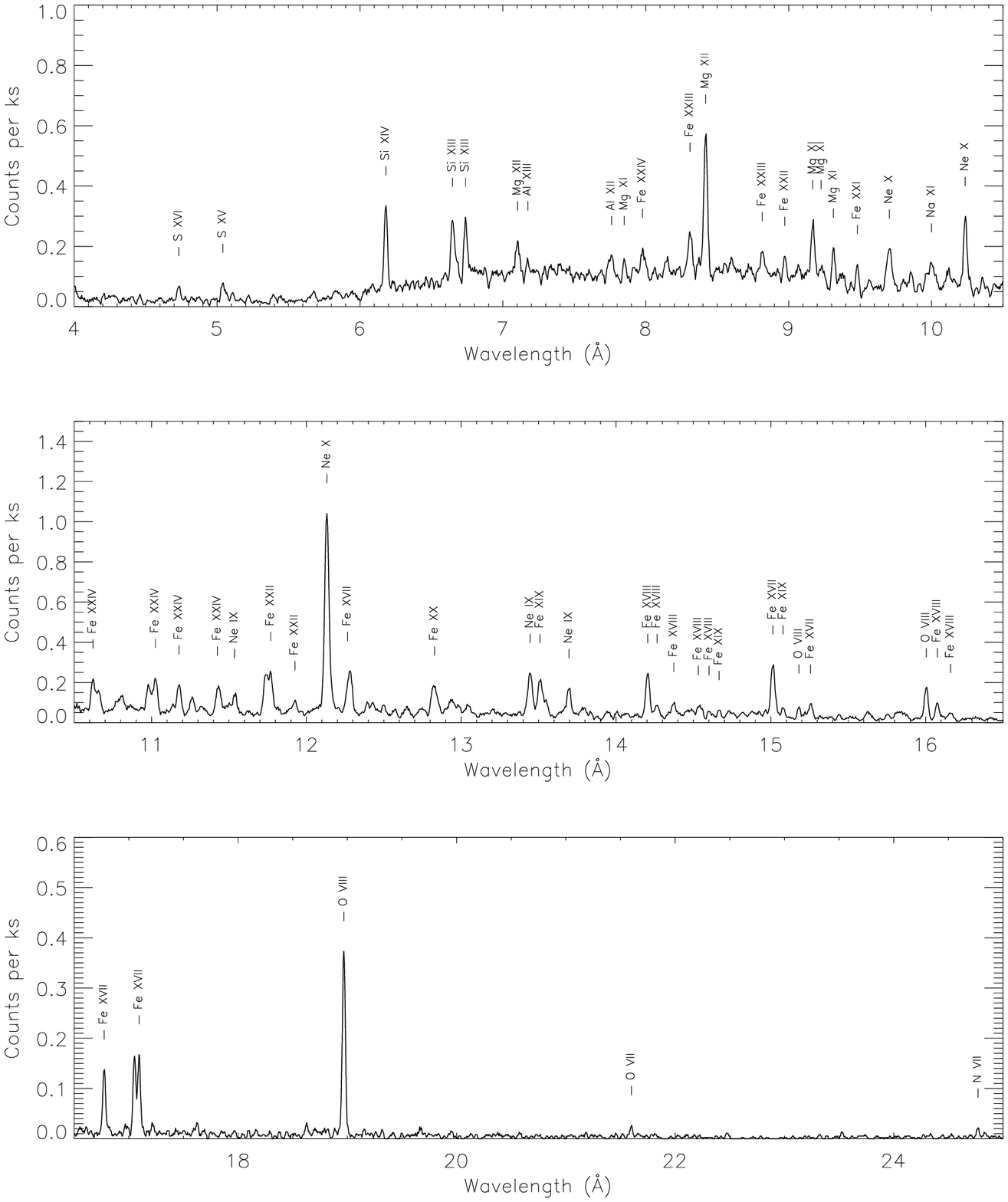}
\caption{PZ~Tel MEG smoothed spectrum (summed positive and negative orders) with bin size of 0.005\,\AA.}
\label{fig:pztel_MEGspec}
\end{figure*}

In Fig.~\ref{fig:pztel_MEGspec} we illustrate the first order MEG spectrum; labels identify the strongest emission lines. The X-ray luminosity obtained from the first order dispersed events in the interval 6-20\,\AA~is $L_{\rm X}=2.2\times10^{30}\,{\rm erg\,s^{-1}}$, from which we derive $\log (L _{\rm X}/L_{\rm bol})=-3.2$, having adopted $L_{\rm bol}=3.6\times10^{33}\,{\rm erg\,s^{-1}}$ \citep[obtained from the $m_{\rm V}$ and $\bv$ Hipparcos data and the bolometric correction of][]{Flower1996}.

\begin{deluxetable}{lllllcrclrc}
\tabletypesize{\scriptsize}
\tablewidth{18cm}
\tablecaption{MEG Spectrum of PZ Tel. \label{tab:lines_res}}
\tablehead{
\colhead{Label} & \colhead{$\lambda_{\rm obs}$\tablenotemark{a}} & \colhead{$\lambda_{\rm pred}$\tablenotemark{a}} & \colhead{Ion} & \colhead{Transition $(upper \rightarrow lower)$} & \colhead{$\log T_{\rm max}$\tablenotemark{b}} & \colhead{$(CR$} & \colhead{$\pm$} & \colhead{$\sigma\;)$\tablenotemark{c}} & \colhead{$EA$\tablenotemark{d}} & \colhead{$em(T)$\tablenotemark{e}} 
}
\startdata
 1a &     4.73 &     4.73 & \ion{S }{    16} &                                            $\;2p^{}\;^{2}P_{3/2}\; \rightarrow \;1s^{}\;^{2}S_{1/2}$ & 7.40 &     0.30 & $\pm$ & 0.13 &  37.5 &    1,2 \\*
 1b & \nodata  &     4.73 & \ion{S }{    16} &                                            $\;2p^{}\;^{2}P_{1/2}\; \rightarrow \;1s^{}\;^{2}S_{1/2}$ & 7.40 &  \nodata &       &      &       &        \\ 
  2 &     5.04 &     5.04 & \ion{S }{    15} &                                           $\;1s\;2p^{}\;^{1}P_{1}\; \rightarrow \;1s^{2}\;^{1}S_{0}$ & 7.20 &     0.25 & $\pm$ & 0.12 &  35.2 &    1,2 \\ 
 3a &     6.18 &     6.18 & \ion{Si}{    14} &                                            $\;2p^{}\;^{2}P_{3/2}\; \rightarrow \;1s^{}\;^{2}S_{1/2}$ & 7.20 &     1.56 & $\pm$ & 0.22 &  99.7 &    1,2 \\*
 3b & \nodata  &     6.19 & \ion{Si}{    14} &                                            $\;2p^{}\;^{2}P_{1/2}\; \rightarrow \;1s^{}\;^{2}S_{1/2}$ & 7.20 &  \nodata &       &      &       &        \\ 
  4 &     6.65 &     6.65 & \ion{Si}{    13} &                                           $\;1s\;2p^{}\;^{1}P_{1}\; \rightarrow \;1s^{2}\;^{1}S_{0}$ & 7.00 &     1.20 & $\pm$ & 0.22 & 117.4 &    1,2 \\ 
 5a &     6.68 &     6.69 & \ion{Si}{    13} &                                           $\;1s\;2p^{}\;^{3}P_{2}\; \rightarrow \;1s^{2}\;^{1}S_{0}$ & 6.95 &     0.31 & $\pm$ & 0.17 & 118.1 &        \\*
 5b & \nodata  &     6.69 & \ion{Si}{    13} &                                           $\;1s\;2p^{}\;^{3}P_{1}\; \rightarrow \;1s^{2}\;^{1}S_{0}$ & 6.95 &  \nodata &       &      &       &        \\ 
  6 &     6.74 &     6.74 & \ion{Si}{    13} &                                           $\;1s\;2s^{}\;^{3}S_{1}\; \rightarrow \;1s^{2}\;^{1}S_{0}$ & 7.00 &     1.14 & $\pm$ & 0.22 & 131.4 &        \\ 
 7a &     7.10 &     7.11 & \ion{Mg}{    12} &                                            $\;3p^{}\;^{2}P_{3/2}\; \rightarrow \;1s^{}\;^{2}S_{1/2}$ & 7.00 &     0.69 & $\pm$ & 0.19 & 145.5 &        \\*
 7b & \nodata  &     7.11 & \ion{Mg}{    12} &                                            $\;3p^{}\;^{2}P_{1/2}\; \rightarrow \;1s^{}\;^{2}S_{1/2}$ & 7.00 &  \nodata &       &      &       &        \\ 
 8a &     7.18 &     7.17 & \ion{Fe}{    24} &                            $\;1s^{2}\;5p^{}\;^{2}P_{3/2}\; \rightarrow \;1s^{2}\;2s^{}\;^{2}S_{1/2}$ & 7.30 &     0.21 & $\pm$ & 0.17 & 134.3 &    1,2 \\*
 8b & \nodata  &     7.17 & \ion{Al}{    13} &                                            $\;2p^{}\;^{2}P_{3/2}\; \rightarrow \;1s^{}\;^{2}S_{1/2}$ & 7.10 &  \nodata &       &      &       &        \\*
 8c & \nodata  &     7.18 & \ion{Al}{    13} &                                            $\;2p^{}\;^{2}P_{1/2}\; \rightarrow \;1s^{}\;^{2}S_{1/2}$ & 7.10 &  \nodata &       &      &       &        \\ 
  9 &     7.76 &     7.76 & \ion{Al}{    12} &                                           $\;1s\;2p^{}\;^{1}P_{1}\; \rightarrow \;1s^{2}\;^{1}S_{0}$ & 6.90 &     0.44 & $\pm$ & 0.18 & 132.9 &    1,2 \\ 
 10 &     7.85 &     7.85 & \ion{Mg}{    11} &                                           $\;1s\;3p^{}\;^{1}P_{1}\; \rightarrow \;1s^{2}\;^{1}S_{0}$ & 6.80 &     0.23 & $\pm$ & 0.17 & 133.8 &        \\ 
11a &     7.98 &     7.98 & \ion{Fe}{    24} &                            $\;1s^{2}\;4p^{}\;^{2}P_{3/2}\; \rightarrow \;1s^{2}\;2s^{}\;^{2}S_{1/2}$ & 7.30 &     0.44 & $\pm$ & 0.18 & 131.4 &      1 \\*
11b & \nodata  &     7.99 & \ion{Fe}{    24} &                            $\;1s^{2}\;4p^{}\;^{2}P_{1/2}\; \rightarrow \;1s^{2}\;2s^{}\;^{2}S_{1/2}$ & 7.30 &  \nodata &       &      &       &        \\ 
12a &     8.31 &     8.30 & \ion{Fe}{    23} &                                           $\;2s\;4p^{}\;^{1}P_{1}\; \rightarrow \;2s^{2}\;^{1}S_{0}$ & 7.20 &     0.93 & $\pm$ & 0.21 & 134.1 &        \\*
12b & \nodata  &     8.32 & \ion{Fe}{    24} &                            $\;1s^{2}\;4d^{}\;^{2}D_{5/2}\; \rightarrow \;1s^{2}\;2p^{}\;^{2}P_{3/2}$ & 7.30 &  \nodata &       &      &       &        \\ 
13a &     8.42 &     8.42 & \ion{Mg}{    12} &                                            $\;2p^{}\;^{2}P_{3/2}\; \rightarrow \;1s^{}\;^{2}S_{1/2}$ & 7.00 &     3.01 & $\pm$ & 0.29 & 136.7 &    1,2 \\*
13b & \nodata  &     8.42 & \ion{Mg}{    12} &                                            $\;2p^{}\;^{2}P_{1/2}\; \rightarrow \;1s^{}\;^{2}S_{1/2}$ & 7.00 &  \nodata &       &      &       &        \\ 
 14 &     8.82 &     8.81 & \ion{Fe}{    23} &                                        $\;2s\;4d^{}\;^{1}D_{2}\; \rightarrow \;2s\;2p^{}\;^{1}P_{1}$ & 7.20 &     0.54 & $\pm$ & 0.19 & 125.1 &        \\ 
 15 &     8.97 &     8.98 & \ion{Fe}{    22} &                   $\;2s^{2}\;(^{1}S)\;4d^{}\;^{2}D_{3/2}\; \rightarrow \;2s^{2}\;2p^{}\;^{2}P_{1/2}$ & 7.10 &     0.35 & $\pm$ & 0.18 & 120.5 &        \\ 
 16 &     9.17 &     9.17 & \ion{Mg}{    11} &                                           $\;1s\;2p^{}\;^{1}P_{1}\; \rightarrow \;1s^{2}\;^{1}S_{0}$ & 6.80 &     1.42 & $\pm$ & 0.21 &  88.7 &    1,2 \\ 
17a &     9.23 &     9.23 & \ion{Mg}{    11} &                                           $\;1s\;2p^{}\;^{3}P_{2}\; \rightarrow \;1s^{2}\;^{1}S_{0}$ & 6.80 &     0.42 & $\pm$ & 0.16 &  81.6 &        \\*
17b & \nodata  &     9.23 & \ion{Mg}{    11} &                                           $\;1s\;2p^{}\;^{3}P_{1}\; \rightarrow \;1s^{2}\;^{1}S_{0}$ & 6.80 &  \nodata &       &      &       &        \\ 
 18 &     9.31 &     9.31 & \ion{Mg}{    11} &                                           $\;1s\;2s^{}\;^{3}S_{1}\; \rightarrow \;1s^{2}\;^{1}S_{0}$ & 6.80 &     0.69 & $\pm$ & 0.18 &  77.1 &        \\ 
19a &     9.48 &     9.48 & \ion{Fe}{    21} &                           $\;2s^{2}\;2p\;4d^{}\;^{3}D_{1}\; \rightarrow \;2s^{2}\;2p^{2}\;^{3}P_{0}$ & 7.05 &     0.27 & $\pm$ & 0.16 &  69.4 &        \\*
19b & \nodata  &     9.48 & \ion{Ne}{    10} &                                            $\;5p^{}\;^{2}P_{3/2}\; \rightarrow \;1s^{}\;^{2}S_{1/2}$ & 6.80 &  \nodata &       &      &       &        \\*
19c & \nodata  &     9.48 & \ion{Ne}{    10} &                                            $\;5p^{}\;^{2}P_{1/2}\; \rightarrow \;1s^{}\;^{2}S_{1/2}$ & 6.80 &  \nodata &       &      &       &        \\ 
20a &     9.70 &     9.71 & \ion{Ne}{    10} &                                            $\;4p^{}\;^{2}P_{3/2}\; \rightarrow \;1s^{}\;^{2}S_{1/2}$ & 6.80 &     0.99 & $\pm$ & 0.19 &  85.9 &      1 \\*
20b & \nodata  &     9.71 & \ion{Ne}{    10} &                                            $\;4p^{}\;^{2}P_{1/2}\; \rightarrow \;1s^{}\;^{2}S_{1/2}$ & 6.80 &  \nodata &       &      &       &        \\ 
21a &     9.96 &     9.97 & \ion{Ni}{    25} &                                        $\;2s\;3d^{}\;^{1}D_{2}\; \rightarrow \;2s\;2p^{}\;^{1}P_{1}$ & 7.30 &     0.21 & $\pm$ & 0.15 &  84.0 &      1 \\*
21b & \nodata  &     9.98 & \ion{Ni}{    19} &                                       $\;2p^{5}\;4d^{}\;^{1}P_{1}\; \rightarrow \;2p^{6}\;^{1}S_{0}$ & 6.90 &  \nodata &       &      &       &        \\ 
22a &    10.02 &    10.02 & \ion{Na}{    11} &                                            $\;2p^{}\;^{2}P_{3/2}\; \rightarrow \;1s^{}\;^{2}S_{1/2}$ & 6.90 &     0.61 & $\pm$ & 0.17 &  86.5 &      1 \\*
22b & \nodata  &    10.03 & \ion{Na}{    11} &                                            $\;2p^{}\;^{2}P_{1/2}\; \rightarrow \;1s^{}\;^{2}S_{1/2}$ & 6.90 &  \nodata &       &      &       &        \\ 
23a &    10.24 &    10.24 & \ion{Ne}{    10} &                                            $\;3p^{}\;^{2}P_{3/2}\; \rightarrow \;1s^{}\;^{2}S_{1/2}$ & 6.80 &     1.36 & $\pm$ & 0.21 &  81.6 &      1 \\*
23b & \nodata  &    10.24 & \ion{Ne}{    10} &                                            $\;3p^{}\;^{2}P_{1/2}\; \rightarrow \;1s^{}\;^{2}S_{1/2}$ & 6.80 &  \nodata &       &      &       &        \\ 
 24 &    10.62 &    10.62 & \ion{Fe}{    24} &                            $\;1s^{2}\;3p^{}\;^{2}P_{3/2}\; \rightarrow \;1s^{2}\;2s^{}\;^{2}S_{1/2}$ & 7.30 &     0.97 & $\pm$ & 0.19 &  72.1 &      1 \\ 
 25 &    10.66 &    10.66 & \ion{Fe}{    24} &                            $\;1s^{2}\;3p^{}\;^{2}P_{1/2}\; \rightarrow \;1s^{2}\;2s^{}\;^{2}S_{1/2}$ & 7.30 &     0.63 & $\pm$ & 0.18 &  71.4 &      1 \\ 
 26 &    10.98 &    10.98 & \ion{Fe}{    23} &                                           $\;2s\;3p^{}\;^{1}P_{1}\; \rightarrow \;2s^{2}\;^{1}S_{0}$ & 7.20 &     0.83 & $\pm$ & 0.18 &  63.4 &      1 \\ 
27a &    11.02 &    11.02 & \ion{Fe}{    23} &                                           $\;2s\;3p^{}\;^{3}P_{1}\; \rightarrow \;2s^{2}\;^{1}S_{0}$ & 7.20 &     1.06 & $\pm$ & 0.19 &  62.5 &      1 \\*
27b & \nodata  &    11.03 & \ion{Fe}{    24} &                            $\;1s^{2}\;3d^{}\;^{2}D_{3/2}\; \rightarrow \;1s^{2}\;2p^{}\;^{2}P_{1/2}$ & 7.30 &  \nodata &       &      &       &        \\ 
28a &    11.18 &    11.17 & \ion{Fe}{    24} &                            $\;1s^{2}\;3d^{}\;^{2}D_{5/2}\; \rightarrow \;1s^{2}\;2p^{}\;^{2}P_{3/2}$ & 7.30 &     0.87 & $\pm$ & 0.19 &  56.9 &      1 \\*
28b & \nodata  &    11.19 & \ion{Fe}{    24} &                            $\;1s^{2}\;3d^{}\;^{2}D_{3/2}\; \rightarrow \;1s^{2}\;2p^{}\;^{2}P_{3/2}$ & 7.30 &  \nodata &       &      &       &        \\ 
 29 &    11.43 &    11.43 & \ion{Fe}{    24} &                            $\;1s^{2}\;3s^{}\;^{2}S_{1/2}\; \rightarrow \;1s^{2}\;2p^{}\;^{2}P_{3/2}$ & 7.30 &     0.59 & $\pm$ & 0.18 &  55.5 &      1 \\ 
30a &    11.44 &    11.44 & \ion{Fe}{    23} &                                        $\;2s\;3d^{}\;^{3}D_{3}\; \rightarrow \;2s\;2p^{}\;^{3}P_{2}$ & 7.15 &     0.40 & $\pm$ & 0.17 &  55.2 &      1 \\*
30b & \nodata  &    11.44 & \ion{Fe}{    22} &                   $\;2s\;2p\;(^{3}P)\;3p^{}\;^{4}S_{3/2}\; \rightarrow \;2s^{2}\;2p^{}\;^{2}P_{1/2}$ & 7.10 &  \nodata &       &      &       &        \\ 
 31 &    11.54 &    11.55 & \ion{Ne}{     9} &                                           $\;1s\;3p^{}\;^{1}P_{1}\; \rightarrow \;1s^{2}\;^{1}S_{0}$ & 6.60 &     0.57 & $\pm$ & 0.16 &  53.5 &      1 \\ 
 32 &    11.74 &    11.74 & \ion{Fe}{    23} &                                        $\;2s\;3d^{}\;^{1}D_{2}\; \rightarrow \;2s\;2p^{}\;^{1}P_{1}$ & 7.20 &     1.26 & $\pm$ & 0.21 &  49.9 &      1 \\ 
 33 &    11.77 &    11.77 & \ion{Fe}{    22} &                   $\;2s^{2}\;(^{1}S)\;3d^{}\;^{2}D_{3/2}\; \rightarrow \;2s^{2}\;2p^{}\;^{2}P_{1/2}$ & 7.10 &     1.28 & $\pm$ & 0.21 &  49.4 &      1 \\ 
34a &    11.93 &    11.92 & \ion{Fe}{    22} &                   $\;2s^{2}\;(^{1}S)\;3d^{}\;^{2}D_{5/2}\; \rightarrow \;2s^{2}\;2p^{}\;^{2}P_{3/2}$ & 7.10 &     0.46 & $\pm$ & 0.15 &  47.0 &      1 \\*
34b & \nodata  &    11.93 & \ion{Fe}{    22} &                   $\;2s^{2}\;(^{1}S)\;3d^{}\;^{2}D_{3/2}\; \rightarrow \;2s^{2}\;2p^{}\;^{2}P_{3/2}$ & 7.10 &  \nodata &       &      &       &        \\*
34c & \nodata  &    11.94 & \ion{Fe}{    21} &                  $\;2s\;2p^{2}\;(^{4}P)\;3p^{}\;^{3}D_{1}\; \rightarrow \;2s^{2}\;2p^{2}\;^{3}P_{0}$ & 7.10 &  \nodata &       &      &       &        \\ 
35a &    12.13 &    12.12 & \ion{Fe}{    17} &                                       $\;2p^{5}\;4d^{}\;^{1}P_{1}\; \rightarrow \;2p^{6}\;^{1}S_{0}$ & 6.80 &     7.13 & $\pm$ & 0.37 &  42.6 &    1,2 \\*
35b & \nodata  &    12.13 & \ion{Ne}{    10} &                                            $\;2p^{}\;^{2}P_{3/2}\; \rightarrow \;1s^{}\;^{2}S_{1/2}$ & 6.80 &  \nodata &       &      &       &        \\*
35c & \nodata  &    12.14 & \ion{Ne}{    10} &                                            $\;2p^{}\;^{2}P_{1/2}\; \rightarrow \;1s^{}\;^{2}S_{1/2}$ & 6.80 &  \nodata &       &      &       &        \\ 
 36 &    12.27 &    12.26 & \ion{Fe}{    17} &                                       $\;2p^{5}\;4d^{}\;^{3}D_{1}\; \rightarrow \;2p^{6}\;^{1}S_{0}$ & 6.80 &     0.74 & $\pm$ & 0.19 &  41.3 &      1 \\ 
 37 &    12.29 &    12.28 & \ion{Fe}{    21} &                           $\;2s^{2}\;2p\;3d^{}\;^{3}D_{1}\; \rightarrow \;2s^{2}\;2p^{2}\;^{3}P_{0}$ & 7.00 &     1.06 & $\pm$ & 0.20 &  41.1 &    1,2 \\ 
38a &    12.83 &    12.82 & \ion{Fe}{    20} &          $\;2s^{2}\;2p^{2}\;(^{3}P)\;3d^{}\;^{4}P_{5/2}\; \rightarrow \;2s^{2}\;2p^{3}\;^{4}S_{3/2}$ & 7.00 &     1.18 & $\pm$ & 0.18 &  35.0 &      1 \\*
38b & \nodata  &    12.82 & \ion{Fe}{    20} &          $\;2s^{2}\;2p^{2}\;(^{3}P)\;3d^{}\;^{4}P_{3/2}\; \rightarrow \;2s^{2}\;2p^{3}\;^{4}S_{3/2}$ & 7.00 &  \nodata &       &      &       &        \\*
38c & \nodata  &    12.82 & \ion{Fe}{    21} &                      $\;2s\;2p^{2}\;(^{4}P)\;3d^{}\;^{3}P_{2}\; \rightarrow \;2s\;2p^{3}\;^{3}D_{1}$ & 7.00 &  \nodata &       &      &       &        \\*
38d & \nodata  &    12.83 & \ion{Fe}{    20} &          $\;2s^{2}\;2p^{2}\;(^{3}P)\;3d^{}\;^{4}P_{1/2}\; \rightarrow \;2s^{2}\;2p^{3}\;^{4}S_{3/2}$ & 7.00 &  \nodata &       &      &       &        \\ 
39a &    13.45 &    13.45 & \ion{Ne}{     9} &                                           $\;1s\;2p^{}\;^{1}P_{1}\; \rightarrow \;1s^{2}\;^{1}S_{0}$ & 6.60 &     1.74 & $\pm$ & 0.24 &  28.4 &    1,2 \\*
39b & \nodata  &    13.46 & \ion{Fe}{    19} &                      $\;2p^{3}\;(^{2}D)\;3d^{}\;^{3}D_{1}\; \rightarrow \;2s^{2}\;2p^{4}\;^{3}P_{0}$ & 6.90 &  \nodata &       &      &       &        \\*
39c & \nodata  &    13.46 & \ion{Fe}{    19} &                      $\;2p^{3}\;(^{2}D)\;3d^{}\;^{3}S_{1}\; \rightarrow \;2s^{2}\;2p^{4}\;^{3}P_{2}$ & 6.90 &  \nodata &       &      &       &        \\ 
40a &    13.51 &    13.51 & \ion{Fe}{    21} &                      $\;2s\;2p^{2}\;(^{4}P)\;3s^{}\;^{3}P_{0}\; \rightarrow \;2s\;2p^{3}\;^{3}D_{1}$ & 7.00 &     1.30 & $\pm$ & 0.19 &  27.8 &      1 \\*
40b & \nodata  &    13.52 & \ion{Fe}{    19} &                      $\;2p^{3}\;(^{2}D)\;3d^{}\;^{3}D_{3}\; \rightarrow \;2s^{2}\;2p^{4}\;^{3}P_{2}$ & 6.90 &  \nodata &       &      &       &        \\*
40c & \nodata  &    13.53 & \ion{Fe}{    19} &                      $\;2p^{3}\;(^{2}D)\;3d^{}\;^{3}D_{2}\; \rightarrow \;2s^{2}\;2p^{4}\;^{3}P_{2}$ & 6.90 &  \nodata &       &      &       &        \\ 
41a &    13.55 &    13.55 & \ion{Fe}{    19} &                      $\;2p^{3}\;(^{2}P)\;3d^{}\;^{3}D_{2}\; \rightarrow \;2s^{2}\;2p^{4}\;^{3}P_{1}$ & 6.90 &     0.46 & $\pm$ & 0.15 &  27.4 &        \\*
41b & \nodata  &    13.55 & \ion{Ne}{     9} &                                           $\;1s\;2p^{}\;^{3}P_{1}\; \rightarrow \;1s^{2}\;^{1}S_{0}$ & 6.55 &  \nodata &       &      &       &        \\*
41c & \nodata  &    13.57 & \ion{Fe}{    19} &                      $\;2p^{3}\;(^{2}D)\;3d^{}\;^{3}P_{2}\; \rightarrow \;2s^{2}\;2p^{4}\;^{3}P_{2}$ & 6.90 &  \nodata &       &      &       &        \\*
41d & \nodata  &    13.57 & \ion{Fe}{    21} &                               $\;2s^{2}\;2p\;3p^{}\;^{1}S_{0}\; \rightarrow \;2s\;2p^{3}\;^{3}D_{1}$ & 7.00 &  \nodata &       &      &       &        \\ 
42a &    13.70 &    13.70 & \ion{Ne}{     9} &                                           $\;1s\;2s^{}\;^{3}S_{1}\; \rightarrow \;1s^{2}\;^{1}S_{0}$ & 6.60 &     0.98 & $\pm$ & 0.18 &  23.9 &        \\*
42b & \nodata  &    13.70 & \ion{Fe}{    19} &                      $\;2p^{3}\;(^{2}D)\;3d^{}\;^{3}D_{2}\; \rightarrow \;2s^{2}\;2p^{4}\;^{3}P_{1}$ & 6.90 &  \nodata &       &      &       &        \\*
42c & \nodata  &    13.70 & \ion{Fe}{    19} &                      $\;2p^{3}\;(^{2}P)\;3d^{}\;^{3}D_{2}\; \rightarrow \;2s^{2}\;2p^{4}\;^{1}D_{2}$ & 6.90 &  \nodata &       &      &       &        \\ 
43a &    14.21 &    14.20 & \ion{Fe}{    18} &                  $\;2p^{4}\;(^{1}D)\;3d^{}\;^{2}D_{5/2}\; \rightarrow \;2s^{2}\;2p^{5}\;^{2}P_{3/2}$ & 6.90 &     1.51 & $\pm$ & 0.20 &  22.7 &    1,2 \\*
43b & \nodata  &    14.21 & \ion{Fe}{    18} &                  $\;2p^{4}\;(^{1}D)\;3d^{}\;^{2}P_{3/2}\; \rightarrow \;2s^{2}\;2p^{5}\;^{2}P_{3/2}$ & 6.90 &  \nodata &       &      &       &        \\ 
44a &    14.27 &    14.26 & \ion{Fe}{    18} &                  $\;2p^{4}\;(^{1}D)\;3d^{}\;^{2}S_{1/2}\; \rightarrow \;2s^{2}\;2p^{5}\;^{2}P_{3/2}$ & 6.90 &     0.37 & $\pm$ & 0.14 &  25.1 &      1 \\*
44b & \nodata  &    14.27 & \ion{Fe}{    18} &                  $\;2p^{4}\;(^{1}D)\;3d^{}\;^{2}F_{5/2}\; \rightarrow \;2s^{2}\;2p^{5}\;^{2}P_{3/2}$ & 6.90 &  \nodata &       &      &       &        \\*
44c & \nodata  &    14.27 & \ion{Fe}{    20} &                  $\;2s\;2p^{3}\;(^{5}S)\;3s^{}\;^{4}S_{3/2}\; \rightarrow \;2s\;2p^{4}\;^{4}P_{5/2}$ & 7.00 &  \nodata &       &      &       &        \\ 
45a &    14.37 &    14.36 & \ion{Fe}{    18} &                  $\;2p^{4}\;(^{1}D)\;3d^{}\;^{2}D_{3/2}\; \rightarrow \;2s^{2}\;2p^{5}\;^{2}P_{1/2}$ & 6.90 &     0.52 & $\pm$ & 0.15 &  25.3 &      1 \\*
45b & \nodata  &    14.37 & \ion{Fe}{    18} &                  $\;2p^{4}\;(^{3}P)\;3d^{}\;^{2}D_{5/2}\; \rightarrow \;2s^{2}\;2p^{5}\;^{2}P_{3/2}$ & 6.90 &  \nodata &       &      &       &        \\ 
 46 &    14.53 &    14.53 & \ion{Fe}{    18} &                  $\;2p^{4}\;(^{3}P)\;3d^{}\;^{2}F_{5/2}\; \rightarrow \;2s^{2}\;2p^{5}\;^{2}P_{3/2}$ & 6.90 &     0.33 & $\pm$ & 0.13 &  24.2 &      1 \\ 
 47 &    14.55 &    14.55 & \ion{Fe}{    18} &                  $\;2p^{4}\;(^{3}P)\;3d^{}\;^{4}P_{3/2}\; \rightarrow \;2s^{2}\;2p^{5}\;^{2}P_{3/2}$ & 6.90 &     0.18 & $\pm$ & 0.13 &  24.0 &      1 \\ 
48a &    14.60 &    14.58 & \ion{Fe}{    18} &                  $\;2p^{4}\;(^{3}P)\;3d^{}\;^{4}P_{1/2}\; \rightarrow \;2s^{2}\;2p^{5}\;^{2}P_{3/2}$ & 6.90 &     0.19 & $\pm$ & 0.12 &  23.4 &        \\*
48b & \nodata  &    14.61 & \ion{Fe}{    18} &                  $\;2p^{4}\;(^{3}P)\;3d^{}\;^{2}P_{3/2}\; \rightarrow \;2s^{2}\;2p^{5}\;^{2}P_{1/2}$ & 6.90 &  \nodata &       &      &       &        \\ 
49a &    14.67 &    14.67 & \ion{Fe}{    19} &                      $\;2p^{3}\;(^{2}D)\;3s^{}\;^{3}D_{3}\; \rightarrow \;2s^{2}\;2p^{4}\;^{3}P_{2}$ & 6.90 &     0.27 & $\pm$ & 0.13 &  23.7 &      1 \\*
49b & \nodata  &    14.67 & \ion{Fe}{    18} &                  $\;2p^{4}\;(^{3}P)\;3d^{}\;^{2}D_{3/2}\; \rightarrow \;2s^{2}\;2p^{5}\;^{2}P_{1/2}$ & 6.90 &  \nodata &       &      &       &        \\ 
 50 &    15.01 &    15.02 & \ion{Fe}{    17} &                                       $\;2p^{5}\;3d^{}\;^{1}P_{1}\; \rightarrow \;2p^{6}\;^{1}S_{0}$ & 6.75 &     1.85 & $\pm$ & 0.23 &  19.4 &    1,2 \\ 
 51 &    15.08 &    15.08 & \ion{Fe}{    19} &                      $\;2s\;2p^{4}\;(^{4}P)\;3s^{}\;^{3}P_{2}\; \rightarrow \;2s\;2p^{5}\;^{3}P_{2}$ & 6.90 &     0.31 & $\pm$ & 0.13 &  20.0 &      1 \\ 
52a &    15.18 &    15.18 & \ion{O }{     8} &                                            $\;4p^{}\;^{2}P_{3/2}\; \rightarrow \;1s^{}\;^{2}S_{1/2}$ & 6.50 &     0.29 & $\pm$ & 0.13 &  19.5 &      1 \\*
52b & \nodata  &    15.18 & \ion{O }{     8} &                                            $\;4p^{}\;^{2}P_{1/2}\; \rightarrow \;1s^{}\;^{2}S_{1/2}$ & 6.50 &  \nodata &       &      &       &        \\*
52c & \nodata  &    15.20 & \ion{Fe}{    19} &                      $\;2p^{3}\;(^{4}S)\;3s^{}\;^{5}S_{2}\; \rightarrow \;2s^{2}\;2p^{4}\;^{3}P_{2}$ & 6.90 &  \nodata &       &      &       &        \\ 
 53 &    15.26 &    15.26 & \ion{Fe}{    17} &                                       $\;2p^{5}\;3d^{}\;^{3}D_{1}\; \rightarrow \;2p^{6}\;^{1}S_{0}$ & 6.75 &     0.44 & $\pm$ & 0.14 &  18.1 &      1 \\ 
54a &    16.00 &    16.00 & \ion{Fe}{    18} &                  $\;2p^{4}\;(^{3}P)\;3s^{}\;^{2}P_{3/2}\; \rightarrow \;2s^{2}\;2p^{5}\;^{2}P_{3/2}$ & 6.90 &     1.00 & $\pm$ & 0.17 &  16.4 &      1 \\*
54b & \nodata  &    16.01 & \ion{O }{     8} &                                            $\;3p^{}\;^{2}P_{3/2}\; \rightarrow \;1s^{}\;^{2}S_{1/2}$ & 6.50 &  \nodata &       &      &       &        \\*
54c & \nodata  &    16.01 & \ion{O }{     8} &                                            $\;3p^{}\;^{2}P_{1/2}\; \rightarrow \;1s^{}\;^{2}S_{1/2}$ & 6.50 &  \nodata &       &      &       &        \\ 
 55 &    16.08 &    16.07 & \ion{Fe}{    18} &                  $\;2p^{4}\;(^{3}P)\;3s^{}\;^{4}P_{5/2}\; \rightarrow \;2s^{2}\;2p^{5}\;^{2}P_{3/2}$ & 6.90 &     0.52 & $\pm$ & 0.14 &  16.1 &      1 \\ 
 56 &    16.11 &    16.12 & \ion{Fe}{    19} &                          $\;2p^{3}\;(^{2}D)\;3p^{}\;^{3}P_{2}\; \rightarrow \;2s\;2p^{5}\;^{3}P_{2}$ & 6.90 &     0.04 & $\pm$ & 0.09 &  16.0 &        \\ 
 57 &    16.16 &    16.17 & \ion{Fe}{    18} &                  $\;2s\;2p^{5}\;(^{3}P)\;3s^{}\;^{2}P_{3/2}\; \rightarrow \;2s\;2p^{6}\;^{2}S_{1/2}$ & 6.90 &     0.25 & $\pm$ & 0.12 &  15.7 &      1 \\ 
 58 &    16.77 &    16.78 & \ion{Fe}{    17} &                                       $\;2p^{5}\;3s^{}\;^{3}P_{1}\; \rightarrow \;2p^{6}\;^{1}S_{0}$ & 6.70 &     0.83 & $\pm$ & 0.16 &  12.4 &      1 \\ 
 59 &    17.05 &    17.05 & \ion{Fe}{    17} &                                       $\;2p^{5}\;3s^{}\;^{1}P_{1}\; \rightarrow \;2p^{6}\;^{1}S_{0}$ & 6.70 &     0.96 & $\pm$ & 0.17 &  11.7 &      1 \\ 
 60 &    17.09 &    17.10 & \ion{Fe}{    17} &                                       $\;2p^{5}\;3s^{}\;^{3}P_{2}\; \rightarrow \;2p^{6}\;^{1}S_{0}$ & 6.70 &     0.91 & $\pm$ & 0.17 &  11.1 &      1 \\ 
61a &    18.97 &    18.97 & \ion{O }{     8} &                                            $\;2p^{}\;^{2}P_{3/2}\; \rightarrow \;1s^{}\;^{2}S_{1/2}$ & 6.50 &     2.44 & $\pm$ & 0.24 &   7.2 &    1,2 \\*
61b & \nodata  &    18.97 & \ion{O }{     8} &                                            $\;2p^{}\;^{2}P_{1/2}\; \rightarrow \;1s^{}\;^{2}S_{1/2}$ & 6.50 &  \nodata &       &      &       &        \\ 
 62 &    21.60 &    21.60 & \ion{O }{     7} &                                           $\;1s\;2p^{}\;^{1}P_{1}\; \rightarrow \;1s^{2}\;^{1}S_{0}$ & 6.30 &     0.13 & $\pm$ & 0.10 &   3.3 &    1,2 \\ 
63a &    24.77 &    24.78 & \ion{N }{     7} &                                            $\;2p^{}\;^{2}P_{3/2}\; \rightarrow \;1s^{}\;^{2}S_{1/2}$ & 6.30 &     0.12 & $\pm$ & 0.10 &   3.0 &      1 \\*
63b & \nodata  &    24.78 & \ion{N }{     7} &                                            $\;2p^{}\;^{2}P_{1/2}\; \rightarrow \;1s^{}\;^{2}S_{1/2}$ & 6.30 &  \nodata &       &      &       &        \\ 
\enddata
\tablenotetext{a}{~Observed and predicted (CHIANTI database) wavelengths in \AA. In case of unresolved blends, identified by the same label number, we list the main components in order of increasing predicted wavelength.}
\tablenotetext{b}{~Temperature in K of maximum emissivity.}
\tablenotetext{c}{~Count rates in ${\rm cts\,ks^{-1}}$ with uncertainties at the 68\% confidence level. In the cases of unresolved blends, identified by the same label number, we report only the total line count rates.}
\tablenotetext{d}{~MEG first order effective area in ${\rm cm^{2}}$, needed to convert the measured count rate to flux.}
\tablenotetext{e}{~Lines selected to derive the emission measure distribution with approach 1 or 2.}
\end{deluxetable}

\clearpage

\section{ANALYSIS}
\label{analysis}

The MEG first order spectrum has more than $1.6\times10^{4}$ total counts and contains lines with good signal-to-noise ratios. The HEG spectrum consists of $5.5\times10^{3}$\,cts, with few lines having sufficient counts to provide useful fluxes. We have therefore based our analysis on the MEG spectrum only. The 
MEG background is extremely low and contains an average of only 0.02 counts per ACIS pixel, and it was therefore ignored.

The spectral analysis of PZ~Tel has been performed using the PINTofALE~V1.5 software package \citep{KashyapDrake2000}. We have adopted the CHIANTI~V4 emission line database \citep{YoungDelZanna2003} and the \citet{MazzottaMazzitelli1998} ionization equilibrium.

Spectral line fluxes were obtained by fitting the counts histograms with a modified Lorentzian profile of the type

\begin{displaymath}
I(\lambda)=I_{\rm max}\left[1+\left(\frac{\lambda-\lambda_{0}}
{\Delta\lambda}\right)^{2}\right]^{-\beta}
\end{displaymath}

\noindent with $\beta=2.4$ plus piece-wise constant continuum level. For the continuum placement we have adopted the following procedure. As a first step we have guessed an emission measure distribution $em(T)$ and set of abundances in order to predict a continuum level for measuring line fluxes. From these line fluxes we evaluated the $em(T)$ and abundances for PZ~Tel, and updated the predicted continuum level. Considering this new predicted continuum we have once again measured the line fluxes. We performed several iterations of the above procedure until convergence, i.e. until the continuum used for the line fitting agrees with the continuum predicted on the basis of the reconstructed $em(T)$ and computed abundances.

In Table~\ref{tab:lines_res} we report the strongest lines identified in the PZ~Tel spectrum, together with their measured MEG count rates. As stated in \S~\ref{obs}, significant variability of the X-ray emission occurred during the observation, but the MEG signal to noise ratio did not allow us to perform time-resolved spectral analysis of the data. On the other hand, the integrated PZ~Tel spectrum provides a good representation of the average state of an active corona where a low-level flare-like activity is present at all times.

The hydrogen column density toward PZ~Tel is not known, and there is insufficient information in the HETG spectrum to estimate it. For all the following analyses we have therefore assumed an hydrogen column density of $10^{19}\,{\rm cm^{-2}}$, appropriate for the distance of PZ~Tel (49.7\,pc) and an average interstellar medium density of $0.1\,{\rm cm^{-3}}$. We have estimated that if the hydrogen column density is as high as $10^{20}\,{\rm cm^{-2}}$, the effect would be marginally observable only for the O and N lines at $\lambda>18$\,\AA, with an absorption of about 5-10\%.

\subsection{Emission Measure Distribution and Abundance Reconstruction}
\label{mcmc}

We have derived the emission measure distribution from individual line fluxes: one of the main advantages of this procedure, in comparison with a global fitting approach, is that we can use only the spectral information that we consider most reliable. This choice involves the question of which lines to include in the $em(T)$ reconstruction and which to exclude. In principle one would expect that lines which form over the same temperature range should give consistent emission measure values, and therefore different line selections should not provide significant differences in the final $em(T)$. However, in some cases the above assumption will not be correct because of uncertainties in the atomic data, hidden line blending, and other line measurement difficulties. 

We have explored two different choices of spectral lines, which we describe below. For both of these selections we derived the emission measure distribution using the method proposed by \citet{KashyapDrake1998}, which performs a search in the $em(T)$ parameter space using a Monte Carlo Markov Chain method, with the aim of maximizing the probability of obtaining equal model and observed line fluxes. Moreover, we have also used two different methods for the abundance determination so that the two approaches are effectively independent. One of the aims of this is to compare the results from the different approaches in order to find which results are the least affected by the methods and hence more robust. Hereafter we will refer to the first selection as Approach~1, and to the second as Approach~2.

\subsubsection{Approach 1}
\label{crit1}

The first Approach is based on the use of all lines strong enough to yield a useful flux measurement, and that are also reasonably isolated, density independent and not affected by identification problems. With this choice the $em(T)$ analysis is based on the most reliable line fluxes among those measured. Note that, for a solar abundance corona, this selection criterion will produce a set of measurements mainly composed of iron lines. For the MEG spectrum of PZ~Tel the lines included in this selection are indicated in the last column of Table~\ref{tab:lines_res}.

With this choice we are able to derive, besides the $em(T)$, the abundances, relative to that of iron, of all the chemical elements for which at least one line has been included in the selection. The abundance reconstruction is performed simultaneously with the $em(T)$ as follows: we first used only the selected iron lines to obtain an initial $em(T)$; we then added lines of other elements, one element at a time, in order to obtain a new $em(T)$ and the abundances of the involved elements relative to Fe. At each step we kept the previous set of $em(T)$ and abundances as starting point for the new reconstruction. The sequence of elements has been estabilished in order to gradually extend the temperature range where the $em(T)$ is constrained. Finally, the Fe abundance has been evaluated by comparing the observed continuum with the predictions based on different metallicity values \citep*[see][for more details]{ArgiroffiMaggio2003}.

In principle, this selection criterion allows us to reach the finest temperature resolution in the $em(T)$ because of the large number of lines involved. While many of the lines have emissivity functions that peak at the same $T_{\rm max}$, these emissivity functions can have significantly different shapes and hence can provide more information on the structure of the $em(T)$ than a single line at the same temperature. Of course, this assumption does depend on there not being significant errors in the emissivity functions of the included lines that might instead lead to distortions of the derived $em(T)$ relative to its true form.

\subsubsection{Approach 2}
\label{crit2}

The second Approach is based on the use of lines produced by He-like and H-like ions and has been proposed by \citet{SchmittNess2004} and J.~J.~Drake, in preparation. This choice is motivated by the simplicity of these ions and that they are likely to have more reliable emissivity functions than lines produced by ions with a larger number of electrons, and in particular the Fe lines. Note that by adopting this selection it becomes necessary to separate the $em(T)$ reconstruction from the abundance evaluation because these quantities are more critically correlated than in the case of Approach~1. This is because there is no element, like iron in Approach~1, whose lines cover the whole temperature range. In order to separate the $em(T)$ and abundance reconstruction, we have used the flux {\em ratios} of the He-like triplet resonance line to the H-like Ly$\alpha$ line for each of the different elements for which these lines have been significantly detected. In addition to these line ratios, we have also included three prominent Fe lines for a better sampling of the temperature range. In the case of PZ~Tel the lines selected are marked in the last column of Table~\ref{tab:lines_res}.

In applying this method we also included measurements of the continuum flux at 2.4-3.4\,\AA, 5.1-6.1\,\AA, and 19-20\,\AA. Fluxes were obtained by summing the counts in each interval. We have checked that no strong line features were present in these intervals, in both the observed and predicted spectra. These continuum measurements serve two purposes. Firstly, they provide additional temperature information, based on the fall-off of the dominant bremsstrahlung continuum toward shorter wavelengths. Secondly, they provide a normalization for the $em(T)$ since our approach based on line ratios described above provides only an estimate of the {\em shape} of the function. An absolute measurement of $em(T)$ also enables us to express element abundances relative to hydrogen. Element abundances were determined when a satisfactory $em(T)$ had been derived by comparing predicted and observed line fluxes for each element.

This line ratio method was developed by J.~J.~Drake, in preparation, in order to investigate coronal element abundances in a large sample of stars. The H- and He-like resonance transitions of the abundant elements N, O, Ne, Mg and Si are the brightest lines in coronal spectra and so should be detected in all stars for which well-exposed spectra have been obtained. One advantage of this method is, then, that the same analysis using the same lines can be performed on many stars, limiting any systematic bias caused by star-to-star line selection differences. One disadvantage, at least in principle, is that the line emissivities of He-like and H-like ions have in general a larger width in temperature than those of iron lines, for example. Any fine structure in the $em(T)$ function will therefore be more likely to be lost in this method than in an approach using a more extensive line list, such as our Approach~1.

\begin{figure}[t]
\scalebox{0.45}{
\includegraphics
{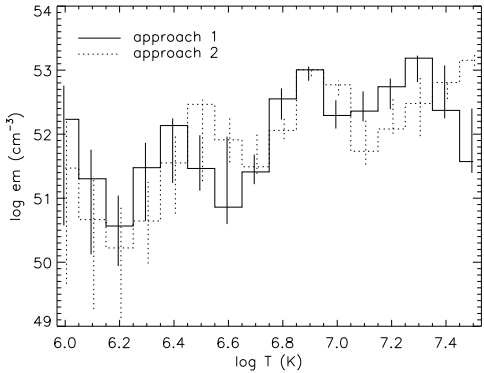}}
\caption{Emission measure distribution $em(T)$ of PZ~Tel derived with the two criteria introduced in \S~\ref{mcmc}.}
\label{fig:pztel_em}
\end{figure}

\begin{figure}[t]
\scalebox{0.45}{
\includegraphics
{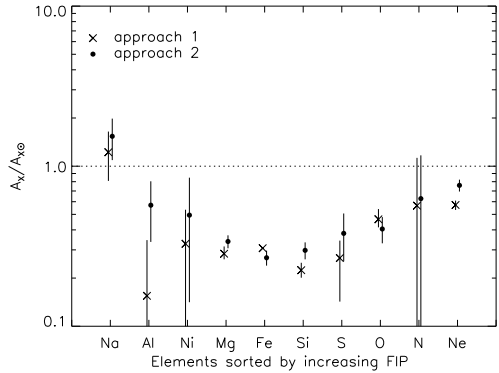}}
\caption{Abundances in solar units \citep{GrevesseSauval1998} vs. elements sorted by increasing First Ionization Potential (FIP) for both the criteria.}
\label{fig:pztel_abund}
\end{figure}

\begin{deluxetable}{lcccc}
\tablecolumns{5}
\tablewidth{0pc}
\tablecaption{Element Abundances $A_{X}=N_{X}/N_{\rm H}$.
\label{tab:tab_abund}}
\tablehead{
\colhead{}      & \multicolumn{2}{c}{Approach 1}  & \multicolumn{2}{c}{Approach 2} \\
\colhead{Elem.} & \colhead{$A_{X}/A_{X\sun}$} & \colhead{$A_{X}$} &
\colhead{$A_{X}/A_{X\sun}$} & \colhead{$A_{X}$}
}
\startdata
Na      & $1.2\pm0.4$   & $(2.6\pm0.9)\times10^{-6}$ & $1.5\pm0.4$   & $(3.3\pm1.0)\times10^{-6}$ \\
Al      & $\le0.3$      & $\le1.0\times10^{-6}$      & $0.6\pm0.3$   & $(1.7\pm0.7)\times10^{-6}$ \\
Ni      & $\le0.5$      & $\le9.5\times10^{-7}$      & $0.5\pm0.4$   & $(9\pm6)\times10^{-7}$ \\
Mg      & $0.28\pm0.03$ & $(1.1\pm0.1)\times10^{-5}$ & $0.34\pm0.03$ & $(1.3\pm0.1)\times10^{-5}$ \\
Fe      & $\sim0.3$     & $\sim9.7\times10^{-6}$     & $0.27\pm0.03$ & $(8.5\pm0.9)\times10^{-6}$ \\
Si      & $0.22\pm0.03$ & $(7.9\pm0.9)\times10^{-6}$ & $0.30\pm0.04$ & $(1.1\pm0.1)\times10^{-5}$ \\
S       & $0.27\pm0.10$ & $(6\pm2)\times10^{-6}$     & $0.38\pm0.13$ & $(8\pm3)\times10^{-6}$ \\
O       & $0.47\pm0.06$ & $(3.1\pm0.4)\times10^{-4}$ & $0.41\pm0.08$ & $(2.7\pm0.5)\times10^{-4}$ \\
N       & $\le1.1$      & $\le9.4\times10^{-5}$      & $\le1.2$      & $\le9.7\times10^{-5}$ \\
Ne      & $0.57\pm0.04$ & $(6.9\pm0.5)\times10^{-5}$ & $0.76\pm0.06$ & $(9.1\pm0.8)\times10^{-5}$ \\
\enddata
\end{deluxetable}

\begin{figure}[h]
\scalebox{0.45}{
\includegraphics
{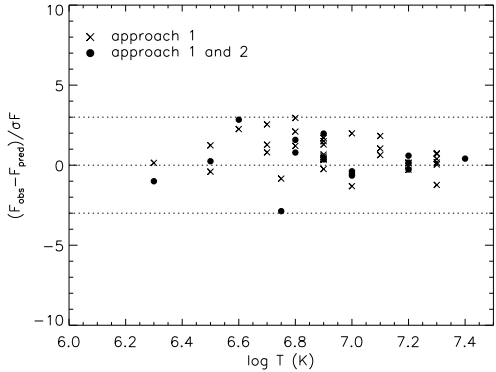}}
\caption{Comparison between observed line fluxes and those predicted by the $em(T)$ obtained with Approach~1. We indicate with the crosses the line fluxes selected for Approach~1 and not for Approach~2, while the filled circles represent the line fluxes included both in Approach~1 and 2.}
\label{fig:check_flx_crit1}
\end{figure}

\begin{figure}[h]
\scalebox{0.45}{
\includegraphics
{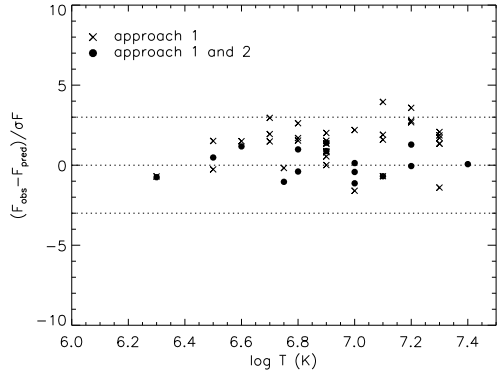}}
\caption{Comparison between observed line fluxes and those predicted by the $em(T)$ obtained with Approach~2. The crosses and filled circles are defined as in Fig~\ref{fig:check_flx_crit1}.}
\label{fig:check_flx_crit2}
\end{figure}

\section{RESULTS}
\label{results}

\subsection{Emission Measure Distribution and Abundances}
\label{emres}

Considering the formation temperatures of the available spectral lines listed in Table~\ref{tab:lines_res}, we have explored the $em(T)$ in the $\log T$ range $6.0-7.5$, with a bin size of $\Delta \log T=0.1$.

In Fig.~\ref{fig:pztel_em} we show the $em(T)$ obtained with the two different methods and line lists, while in Table~\ref{tab:tab_abund} and Fig.~\ref{fig:pztel_abund} we report the derived abundances. As explained in \S~\ref{mcmc}, Approach~1 compares the observed and predicted continuum level in order to evaluate the iron abundance and hence the overall plasma metallicity. Thus we have not assigned an error bar to the Fe abundance in Fig.~\ref{fig:pztel_abund}. This comparison yields an iron abundance of $\sim0.3$ times the currently accepted solar value \citep{GrevesseSauval1998}. Note that the iron abundance so derived plays the role of a global factor both in the $em( T)$ and in the abundance values derived with Approach~1.

The O, Ne, Mg, Al, Si, S, and Fe abundance estimations in Approach~2 were performed using the lines that were used in the $em(T)$ derivation; the N, Na, and Ni abundances were instead evaluated using their measured line counts listed in Table~\ref{tab:lines_res} that were excluded from the $em(T)$ analysis.

The two emission measure distributions obtained with the two approaches (see \S~\ref{crit1} and \ref{crit2}) are compatible within statistical uncertainties in almost all the bins of the explored temperature range. Both $em(T)$ distributions show a maximum at $\log T = 6.9$, which is a common feature for active stars \citep*[e.g.][]{Drake1996,Sanz-ForcadaBrickhouse2003}, and also a significant emission measure for $\log T \ge 7$. However, some differences are present, mainly near $\log T \sim 6.5$ and in the high temperature tails of the distributions.

Several causes might be responsible for these differences. Errors in the atomic data may yield different solutions if different line sets are used. This is one of the primary motives of our Approach~2, which should introduce minimal errors arising from atomic data.

Another problem is due to the fact that the observed line fluxes are related to the $em(T)$ function through a Fredholm equation of first type, which has no unique solution \citep[e.g.][]{CraigBrown1976}. Different $em(T)$ and abundance sets might then reproduce equally well the observed line fluxes within measurement uncertainties. This non-uniqueness might also be affected by the resolution and amplitude of the temperature grid assumed \emph{a priori} for the $em(T)$ reconstruction.

The two different line choices and methods provide consistent abundance values in most cases, indicating that the derived abundances do not depend on small differences in the emission measure distributions. We find a significant discrepancy between the two models only for Si and Ne abundances. These small abundance differences are probably due to the slightly different values of the $em(T)$ obtained with the two approaches in the temperature range where Si and Ne lines are produced.

\begin{deluxetable}{clccc}
\tablecolumns{5}
\tablewidth{0pc}
\tablecaption{Electron Densities Calculated from the He-like Triplets. \label{tab:tab_dens}}
\tablehead{
\colhead{Ion} & \colhead{$\log T_{\rm max}$} & \colhead{$f/i$} & \colhead{$\log N_{\rm e}\,({\rm cm^{-3}})$} 
}
\startdata
\ion{Ne}{9}  & 6.6 & $3.0\pm1.4$ & $<11.8$ \\
\ion{Mg}{11} & 6.8 & $1.7\pm0.8$ & $12.6\pm0.6$ \\
\ion{Si}{13} & 7.0 & $3.5\pm2.1$ & $<13.5$ \\
\enddata
\end{deluxetable}

In Fig.~\ref{fig:check_flx_crit1} and \ref{fig:check_flx_crit2} we show the comparison between observed and predicted line fluxes for both the solutions. From these plots we deduce that, as expected, the $em(T)$ derived with Approach~1 shows a better overall description of the observed line fluxes of PZ~Tel than in the other case, but the predicted line flux values of the resonance line of the He-like ions and Ly$\alpha$ line of the H-like ions are more discrepant than those computed with Approach~2 solution (filled circles in Fig.~\ref{fig:check_flx_crit1} and \ref{fig:check_flx_crit2}). Moreover, we note that the Approach~2 solution describes spectral features formed at lower temperatures somewhat better, as represented by a smaller spread between observed and predicted fluxes for $\log T \le 7.0$. The Approach~1 solution, on the other hand, shows a smaller spread between observed and predicted fluxes for $\log T \ge 7.0$.

Figure~\ref{fig:pztel_abund} shows that the PZ~Tel coronal abundances, relative to solar photospheric values \citep{GrevesseSauval1998}, do not show a strong FIP dependence, but rather show a pattern vs. FIP similar to that of other active stars \citep[e.g.][]{BrinkmanBehar2001,DrakeBrickhouse2001,AudardGudel2003,Sanz-ForcadaMaggio2003}: elements at low and high FIP appear enhanced with respect to the medium FIP elements. It is worth noting that we cannot compute abundance ratios relative to photospheric values because those of PZ~Tel are not known; therefore the \emph{real} abundance vs. FIP pattern may differ significantly from the one plotted in Fig.~\ref{fig:pztel_abund} \citep*[e.g.][]{Sanz-ForcadaFavata2004}.

Finally we note that, using the flux measured for the \ion{Na}{11} Ly$\alpha$ line at 10.02\AA, we have performed --- to our knowledge --- the first coronal Na abundance estimate based on X-ray spectra. The Na abundance is useful for constraining the abundance vs. FIP pattern at low FIP. The identification and subsequent flux estimation of these lines is nontrivial because this spectral region also contains a number of \ion{Fe}{20} and other lines. \citet{AyresBrown2001} and \citet{PhillipsMathioudakis2001} identified the \ion{Na}{11} Ly$\alpha$ line in the HETGS spectra of HR~1099 and Capella, respectively. \citet{Sanz-ForcadaMaggio2003,Sanz-ForcadaFavata2004}, in the HETGS of AB~Dor and $\lambda$~And respectively, measured the flux of a spectral feature at 10.02\,\AA, but they did not identify it because of a lack of Na lines in the Astrophysical Plasma Emission Database (APED). Moreover, in the HETGS spectrum of AR~Lac, analyzed by \citet{HuenemoerderCanizares2003} using the APED database, this emission line at 10.02\,\AA~is clearly visible and does not appear to be easily attributable to other species, such as \ion{Fe}{20}. In conclusion, we suggest that the spectral feature at 10.02\,\AA~observed in several coronal X-ray spectra is to be identified with the \ion{Na}{11} Ly$\alpha$ line, listed in the CHIANTI database.

\subsection{Electron Density}

Estimates of the electron density $N_{\rm e}$ can be obtained using He-like triplets \citep{GabrielJordan1969}. These spectral features consist of three lines (resonance $r$, intercombination $i$ and forbidden $f$) produced by transitions from $n=2$ to $n=1$ levels in He-like ions. For the physical conditions of coronal plasmas, the ratio $f/i$ is predominantly sensitive to density, while the ratio $(i+ f)/r$ is predominantly sensitive to temperature. Note that the density and the temperature inferred from these lines will be weighted averages of for region in which the relevant triplet is formed. 

We have identified and measured in the MEG spectrum of PZ~Tel the He-like triplets of \ion{Ne}{9}, \ion{Mg}{11} and \ion{Si}{13} ions, while for the \ion{O}{7}, \ion{Al}{12} and \ion{S}{15} ions only the
resonance line is measurable (see Table~\ref{tab:lines_res}). In Table~\ref{tab:tab_dens} we report the $f/i$ ratios, together with the derived electron densities $N_{\rm e}$ and $1\sigma$ uncertainties. Note that, because of the unresolved blends of \ion{Ne}{9} $i$ and $f$ lines with \ion{Fe}{19} lines, we have evaluated the \ion{Ne}{9} $f/i$ ratio by subtracting the predicted fluxes of the \ion{Fe}{19} lines from the measured values. We have obtained density values from the \ion{Mg}{11}, while from \ion{Ne}{9} and \ion{Si}{13} we have derived only an upper limit. We also point out here that our measured $f/i$ ratios have uncertainties of more than 40\%\ and are all consistent with their respective low density limits at the $2\sigma$ level.

\section{DISCUSSION AND CONCLUSIONS}
\label{disc}

\begin{figure}[t]
\scalebox{0.45}{
\includegraphics
{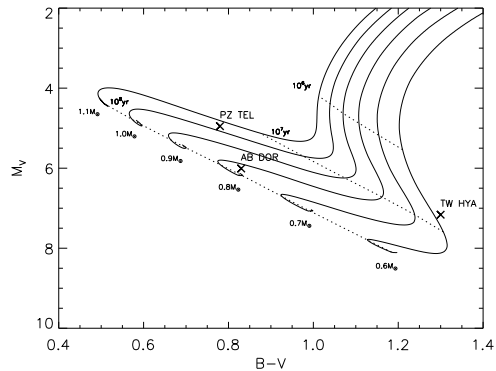}}
\caption{HR diagram for PZ~Tel, TW~Hya and AB~Dor. Solid lines represent the evolutionary tracks of \citet{VenturaZeppieri1998} evaluated for $Z=0.01$. Dotted lines represent the isochrones at $10^6$, $1
0^7$ and $10^{8}$\,yr.}
\label{fig:HR_diagram}
\end{figure}

It is interesting to compare the coronal properties of PZ~Tel found here with those of TW~Hya and AB~Dor, which are slightly younger and older, respectively, than PZ~Tel. TW~Hya is a classical T~Tauri star with a circumstellar disk and an age of about 10\,Myr \citep{WebbZuckerman1999}. AB~Dor is a K1 single star that has almost arrived on the main sequence. \citet{CollierCameronFoing1997} have estimated the age of AB~Dor to be in the range from 20 to 30 \,Myr, while \citet{FavataMicela1998} derived a slightly older age, about $\gtrsim35$\,Myr. In Fig.~\ref{fig:HR_diagram} we show the positions on the HR diagram of these three stars and have also superimposed the evolutionary tracks from \citet{VenturaZeppieri1998} with $Z=0.01$. The values of $M_{V}$ and $\bv$ have been derived from the Hipparcos catalog \citep{PerrymanLindegren1997}. Note however that TW~Hya cannot be placed accurately on the HR diagram because of the large variability observed in its color indices \citep{RucinskiKrautter1983,Mekkaden1998}. \citet{RucinskiKrautter1983} found that the spectrum of TW~Hya is compatible with a K7V star, and we have therefore assumed a $\bv$ color index of 1.30 \citep{Johnson1966}. The positions of these stars on the HR diagram (Fig.~\ref{fig:HR_diagram}) confirm that TW~Hya and AB~Dor are in evolutionary stages just preceding and following the stage of PZ~Tel.

The coronal emission of TW~Hya and AB~Dor has been studied via {\em Chandra} HETGS observations by \citet{KastnerHuenemoerder2002}, \citet{Sanz-ForcadaMaggio2003} and D. Garc{\'{\i}}a-Alvarez, in preparation. Comparison with our present results shows that PZ~Tel is very similar to AB~Dor in its emission measure distribution, element abundances and electron densities. TW~Hya, however, appears to be very different in many respects. Its emission measure distribution peaks at $\log T = 6.5$, and there is no evidence of plasma at $\log T \ge 7.0$ in the observed spectrum, while PZ~Tel and AB~Dor have their maxima at $\log T = 6.9$, and both exhibit considerable emission measure at higher temperatures. There are also abundance differences, especially regarding the Ne/Fe ratio. In TW~Hya, Ne/Fe is larger than the solar value by a factor $\sim10$, whereas for AB~Dor and PZ~Tel the enhancements are less extreme, with factors of $\sim5$ and $\sim2$, respectively\footnote{For the Ne/Fe abundance ratio of PZ~Tel, we have used results obtained with Approach~1. Approach~2 gives a Ne/Fe abundance ratio $\sim 3$, which also tends to confirm the observed trend.}.

\begin{figure}[t]
\scalebox{0.5}{
\includegraphics
{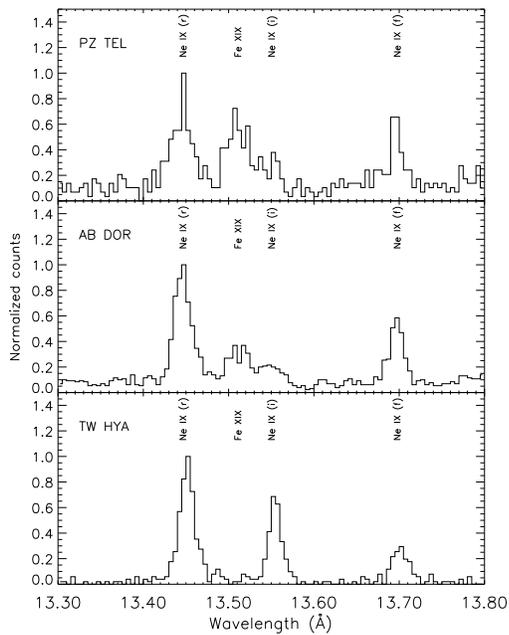}}
\caption{Comparison of the \ion{Ne}{9} spectral region of PZ~Tel, AB~Dor and TW~Hya. For each star we have plotted the first order spectra of {\em Chandra} MEG.}
\label{fig:comp_spec_NeIX}
\end{figure}

Finally, the electron densities of the emitting plasma of TW~Hya, derived from the \ion{O}{7} and \ion{Ne}{9} He-like triplets, turn out to be much higher than those of PZ~Tel and AB~Dor. The differences in densities between PZ~Tel, AB~Dor and TW~Hya can be appreciated in Fig.~\ref{fig:comp_spec_NeIX}, where we show their observed MEG spectra in the region of the \ion{Ne}{9} lines. We show the \ion{Ne}{9} triplet region because it is the only triplet for which the observed spectra have sufficient $S/N$ for all three stars (in fact, the \ion{O}{7} intercombination and forbidden lines in the spectrum of PZ~Tel are not measurable). It is known that the \ion{Ne}{9} triplet is complicated by blends with highly ionized Fe lines that are clearly visible in the PZ~Tel spectrum, e.g., at 13.51\,\AA~(see also Table~\ref{tab:lines_res} for more detail). The \ion{Fe}{19} line at $\sim13.51$\,\AA~in the TW~Hya spectrum is instead not observable because of the peculiar Ne/Fe abundance ratio and the relatively low temperature of the peak in $em(T)$ for this star. However, it is still quite clear that the relative intensities of the $i$ and $f$ lines for PZ~Tel and AB~Dor are very different from those of TW~Hya. In fact the $f/i$ ratio is $2.9\pm0.3$ for AB~Dor, $3.0\pm1.4$ for PZ~Tel and $0.44\pm0.12$ for TW~Hya, implying electron density values of $\log N_{\rm e}\sim 11$, $< 12$, and $\sim 13$, respectively.

As noted in \S~\ref{intro}, \citet{KastnerHuenemoerder2002} attributed the dominance of plasma at a temperature of $3\times 10^6$\,K, with high electron densities of order $\log N_{\rm e}\sim 13$ in TW~Hya, to accretion rather than a magnetically heated corona. In the accretion scenario, the plasma is heated in a shock at the stellar surface. The strong difference we see here between the X-ray emission of the accreting TW~Hya and non-accreting PZ~Tel supports these conclusions. However, there are two outstanding questions that the accretion scenario does not answer: (i) why is the Ne/Fe abundance in the shocked plasma similar to that seen in some coronally active RS~CVn systems (see e.g. \citealt{DrakeBrickhouse2001}; \citealt*{HuenemoerderCanizares2001}) and so different from the presumed value in the accreting circumstellar material? And (ii) why is there very little trace of underlying coronal activity similar to that in PZ~Tel?

If the Ne/Fe abundance seen in TW~Hya is indeed representative of the accreting material, then significant compositional fractionation must be occurring in the circumstellar disk or in the inner disk-magnetosphere region from which the accreting plasma falls onto the star. The lack or weakness of ``normal'' coronal activity on TW~Hya might logically be associated with accretion activity or the presence of substantial disk material close in to the star itself. In this respect, however, it seems that TW~Hya might be unusual among accreting CTTSs. X-ray studies at low resolution of large samples of PMS stars \citep{TsujimotoKoyama2002,NakajimaImanishi2003} have indicated that CTTSs coronal spectra tend to be dominated by plasma at $\log T \sim 7.5$ --- much hotter than in the case of TW~Hya. Note however that CTTSs are often highly absorbed and therefore the high difference in temperature could be the result of a selection bias because of their distances larger than that of TW~Hya.

Our study of PZ~Tel shows it to be much more similar in $em(T)$, abundances and electron densities to AB~Dor than to TW~Hya. Because PZ~Tel is younger than AB~Dor, our results indicate that coronal properties do not change appreciably during the evolution of PMS stars from the post T~Tauri phase to the main sequence. On the other hand, if TW~Hya is assumed to be representative of CTTSs, it follows that as soon as disk dissipation and significant accretion processes have ended, the coronal structure quickly evolves to that of any other star of similar spectral type and rotation rate.

\acknowledgements
CA, AM, GP and SS acknowledge partial support for this work by Agenzia Spaziale Italiana and Ministero dell'Istruzione, Universit\`a e Ricerca. FRH acknowledges partial support from {\it Chandra} grant GO3-4009B, as does CA, during her visit to SAO where much of this work was carried out. JJD was supported by NASA contracts NAS8-39073 and NAS8-03060 to the {\em Chandra} X-ray Center. JJD also thanks the NASA AISRP for providing financial assistance for the development of the PINTofALE package under NASA grant NAG5-9322.

\bibliographystyle{apj}
\bibliography{x}

\end{document}